\documentclass[times]{qjrms4}

\usepackage[colorlinks,bookmarksopen,bookmarksnumbered]{}

\usepackage[draft]{hyperref}

\usepackage{amsmath}
\usepackage{amssymb}
\usepackage{upgreek}
\usepackage{bbm}
\usepackage{dsfont}
\usepackage{sansmath}
\usepackage{gensymb}
\usepackage{mathrsfs}
\usepackage[T1]{fontenc}

\begin{document}

\runningheads{I.~Polichtchouk and T.~G.~Shepherd}{Zonal-mean circulation
  response to reduced air-sea momentum roughness}

\title{Zonal-mean circulation response to reduced
  air-sea momentum roughness}

\author{I.~Polichtchouk\corrauth and T.~G.~Shepherd}

\address{Department of Meteorology, University of Reading, Reading RG6 6BX, UK\\}

\corraddr{Department of Meteorology, University of Reading, Earley
  Gate, PO Box 243, Reading, RG6 6BB, UK. E-mail:
  I.Polichtchouk@reading.ac.uk\\}

\begin{abstract}
The impact of uncertainties in surface layer physics on the
atmospheric general circulation is comparatively unexplored. Here the
sensitivity of the zonal-mean circulation to reduced air-sea momentum
roughness ($Z_{0m}$) at low flow speed is investigated with the
Community Atmosphere Model (CAM3). In an aquaplanet framework with
prescribed sea surface temperatures, the response to reduced $Z_{0m}$
resembles the La~Ni$\tilde{\text{n}}$a minus El~Ni$\tilde{\text{n}}$o
response to El~Ni$\tilde{\text{n}}$o Southern Oscillation variability
with: i) a poleward shift of the mid-latitude westerlies extending all
the way to the surface; ii) a weak poleward shift of the subtropical
descent region; and iii) a weakening of the Hadley circulation, which
is generally also accompanied by a poleward shift of the
inter-tropical convergence zone (ITCZ) and the tropical surface
easterlies.  Mechanism-denial experiments show this response to be
initiated by the reduction of tropical latent and sensible heat
fluxes, effected by reducing $Z_{0m}$. The circulation response is
elucidated by considering the effect of the tropical energy fluxes on
the Hadley circulation strength, the upper tropospheric critical layer
latitudes, and the lower-tropospheric baroclinic eddy forcing. The
ITCZ shift is understood via moist static energy budget analysis in
the tropics. The circulation response to reduced $Z_{0m}$ carries over
to more complex setups with seasonal cycle, full complexity of
atmosphere-ice-land-ocean interaction, and a slab ocean lower boundary
condition. Hence, relatively small changes in the surface
parameterization parameters can lead to a significant circulation
response.

\end{abstract}


\maketitle

\section{Introduction}\label{sec1}

Atmospheric general circulation models (GCMs) exhibit persistent
biases in their representation of the circulation at a quantitative
level. Among such biases are the too equatorward location of the
eddy-driven jet in both hemispheres \citep[e.g.][]{Barnes10,Simpson13}
and the exaggeration of the intertropical convergence zone (ITCZ)
split over the eastern Pacific \citep[e.g.][]{Hwang13}. Moreover,
climate models exhibit a wide range of predicted changes in these
circulation patterns under global warming
\citep[e.g.][]{Stevens13,Shepherd14}. Uncertainties in parameterized
processes are thought to be in part responsible for the persistent
model biases and the divergence in model projections. While the cloud
and convection parameterizations have thus far received the most
attention, the surface processes remain comparatively unexplored.

The interaction of the surface layer with an overlying atmosphere is
an important process governing the atmospheric circulation. Eddy
momentum fluxes maintain the eddy-driven jet against surface
friction. Latent and sensible heat fluxes supply energy to the
overlying atmosphere and thereby influence the thermally direct Hadley
circulation (HC) and the associated ITCZ. Therefore, it is crucial to
accurately represent the surface-atmosphere interaction in any GCM.

Due to the insufficient spatial and temporal resolutions in GCMs, the
explicit representation of the turbulent surface fluxes is not
possible. Hence, the surface fluxes must be parameterized. In
state-of-the-art GCMs, the turbulent fluxes are treated with a bulk
exchange formulation based on the Monin-Obukhov similarity theory
(MO). Because many parameters underlying MO are empirically
determined, uncertainties exist in the surface flux parameterizations.
One such class of empirically determined parameters is surface
exchange coefficients and, in particular, roughness lengths. Exploring
the circulation sensitivity to uncertainties in the roughness length
for momentum is the objective of this study.

The view that a uniform surface layer (i.e., in the absence of
orography) is passive and determined by the eddy momentum fluxes in
the atmosphere has been challenged by several recent studies, which
suggest that the representation of the surface layer can have a direct
impact on the momentum budget and the extra-tropical
circulation. \cite{Chen07} showed that the latitude of the eddy-driven
jet shifts poleward in response to decrease in surface friction in an
idealized dry GCM, in a setup similar to \cite{HeldSuarez94}. In that
setup the surface friction is parameterized as a linear relaxation of
near-surface winds to zero on a specified time scale (i.e. the
Rayleigh drag) and such a parameterization has a direct effect on the
momentum budget only. However, unrealistically large perturbations to
the surface friction are required to produce significant jet
shifts. Moreover, the dry \cite{HeldSuarez94} setup is limited in its
representation of the tropics and hence the effect of reduced surface
friction on the tropical circulation remains to be quantified. Using a
full-complexity GCM, \cite{Garfinkel11, Garfinkel13} found that
increasing surface roughness for momentum at the air-sea interface for
moderate flow speed (4-20\,m\,s$^{-1}$) in NASA's Goddard Earth
Observing System Chemistry-Climate Model (GEOS-5) improved the Southern
Hemisphere stratospheric circulation together with surface winds and
eddy momentum flux convergence aloft over the Southern Ocean.

Details in the surface latent and sensible heat flux
parameterization are also known to affect the tropical
circulation. For example, \cite{Miller92} showed that increasing
latent and sensible heat fluxes at low flow speed (< 5\,m\,s$^{-1}$)
in the ECMWF global forecast model resulted in improvement of seasonal
rainfall distributions, the Indian monsoon circulation, and systematic
tropical model biases. \cite{Numaguti93} found that the combined
structure of the HC and the ITCZ is strongly influenced by the
distribution of the latent heat fluxes in an aquaplanet GCM forced by
zonally- and hemispherically-symmetric SST distributions with a
maximum at the equator. In particular, he showed that despite the
equatorial maximum in SST, a local evaporation minimum can form at the
equator and generate a double ITCZ. This minimum came about from the
near-surface flow speed dependency of the latent heat flux
parameterization (see equation~(\ref{bulks})). If, however, the flow
dependency was removed from the parameterization a single ITCZ formed
at the equator, at the SST maximum.

Similar to \cite{Garfinkel11,Garfinkel13}, the present study explores
the sensitivity of the large-scale circulation to changes in air-sea
momentum roughness (henceforth $Z_{0m}$) within the limits of the
observational constraints. Different from
\cite{Garfinkel11,Garfinkel13}, $Z_{0m}$ is reduced at low flow speed
in the Community Atmosphere Model (CAM3) \citep{Collins06}. Because
reducing $Z_{0m}$ affects both the surface stresses and the latent and
sensible heat fluxes, the large-scale circulation can respond both to
changes in the momentum budget --- as in \cite{Chen07} and
\cite{Garfinkel13} --- and to changes in the thermodynamic budget ---
as in \cite{Miller92} and \cite{Numaguti93}.

By performing a series of simulations with CAM3 under aquaplanet
\citep{Neale00} and AMIP-type \citep{Gates99} setups forced by
prescribed SSTs, it is shown that the zonal-mean circulation response
to reduced $Z_{0m}$ is similar to the La Ni$\tilde{\text{n}}$a minus
El Ni$\tilde{\text{n}}$o response to ENSO variability
\citep[e.g.][]{Lu08,Seager03}, and that the sensitivity is initiated
thermodynamically and from the tropics. The aquaplanet setup is used
to cleanly isolate the circulation response. The setup excludes the
seasonal cycle and the complexity and temporal variability of land,
ocean and ice distribution, but contains the full complexity of the
GCM parameterizations including the MO treatment of the surface
layer. Reducing $Z_{0m}$ at low flow speed reduces the latent and
sensible heat fluxes in the tropics. This cools the tropical
troposphere, weakens the HC and through tropical-extratropical zonally
symmetric teleconnections pushes the mid-latitude jet and the HC
terminus poleward.

As is well known, boundary conditions can alter the admitted
solutions. To determine if the circulation sensitivity to $Z_{0m}$
carries over when the SSTs are free to respond, AMIP-type simulations
with a mixed layer slab ocean lower boundary condition are also
performed. It is shown that the circulation sensitivity is essentially
the same as in the prescribed SST simulations.

The paper is organized as follows. Section~\ref{sec2} briefly reviews
the model, the experimental setup, and the changes made to the surface
layer exchange formulation over the ocean. In section~\ref{sec3},
results from aquaplanet simulations are discussed and the role of
tropical latent and sensible heat fluxes in initiating the circulation
response is isolated. This section also elucidates the circulation
response in the aquaplanet framework via an ensemble of reduced
$Z_{0m}$ switch-on simulations. In section~\ref{sec4}, results from
the AMIP-type simulations with prescribed SSTs and with a slab ocean
lower boundary condition are presented.  Finally, summary and
conclusions are given in section \ref{sec5}.

\section{Method}\label{sec2}
\subsection{Model description}\label{sec2.1}
The GCM used in this study is CAM3, which is the atmospheric component
of the Community Climate System Model (CCSM3). The detailed
description of the pseudospectral dynamical core and the physical
parameterizations of CAM3 can be found in \cite{Collins06}. Moisture
transport is monotonic semi-Lagrangian and time-split in the
horizontal and vertical directions. CAM3 is integrated mostly with T85
truncation. However, the AMIP-type simulations requiring long
integration times are integrated with T42 truncation. The vertical
domain is resolved by 26 levels and the model top is located at
2.917\,hPa. The time step for T85 truncation is $\Delta t=10$~minutes
and the parameterizations are applied over the centred interval of
$2\Delta t=20$~minutes. The $\nabla^4$ constant hyperdiffusion
coefficient is set to $10^{15}$\,m$^4$\,s$^{-1}$ in T85 truncation
simulations. These are the default values used in CAM3.

The surface layer exchange formulation in CAM3 over the ocean follows
\cite{Large82}. Section \ref{sec2.1} discusses the modifications made
to the Large and Pond scheme in the reduced $Z_{0m}$
simulations. Otherwise, all simulations described here use the
standard CAM3 parameterizations: The planetary boundary layer scheme
follows \cite{Holtslag93}; moist convection is parameterized by the
\cite{Zhang95} deep convection scheme; shallow convection is
parameterized by the \cite{Hack94} scheme; the treatments of
microphysics and cloud condensation follow \cite{Boville06}; and the
prognostic cloud water scheme is discussed in \cite{Zhang03}.

\subsection{Changes to the surface exchange formulation}\label{sec2.1}
The bulk formulas used to determine the turbulent momentum (i.e.,
stress $\boldsymbol{\tau}$), latent ($E$), and sensible heat ($H$)
fluxes over the ocean are:
\begin{equation}\label{bulks}
(\boldsymbol {\tau},E,H) = \rho_A |\Delta {\bf v}|(C_D \Delta {\bf v}, C_E
  \Delta q, c_pC_H\Delta \theta),
\end{equation}
where subscript $A$ denotes the lowest model level field; $\rho$ is
density; $|\Delta {\bf v}| = \text{max}(1,\sqrt{u_A^2+v_A^2})$, where
$u$ and $v$ are the zonal and meridional winds in units of m\,s$^{-1}$;
$c_p$ is the (constant) specific heat at constant pressure; $\Delta
{\bf v}=(u_A, v_A)$; $\Delta \theta = \theta_A - T_s$, where $\theta$
is the potential temperature and $T_s$ is the surface temperature;
$\Delta q = q_A - q_s(T_s)$, where $q$ is the specific humidity and
$q_s(T_s)$ is the saturation specific humidity at $T_s$; and
$C_{\{D,E,H\}}$ are the transfer coefficients at the air-sea interface
for $\boldsymbol{\tau}$, $E$, and $H$, respectively. The transfer
coefficients are computed at $10$\,m height and are functions of
stability $\zeta$ and roughness lengths for momentum, evaporation and
sensible heat $Z_{0\{m,e,h\}}$, such that
$C_{\{D,E,H\}}=C_{\{D,E,H\}}(\zeta, Z_{0\{m,e,h\}})$. For neutral
conditions,
\begin{equation}\label{transfer1}
C_{DN}=\frac{\kappa^2}{\ln \left(\frac{10}{Z_{0m}}\right)^2},
\end{equation}
\begin{equation}\label{transfer2}
C_{\{EN,HN\}}=\sqrt{C_{DN}}\frac{\kappa}{\ln\left(\frac{10}{Z_{0\{e,h\}}}\right)},
\end{equation}
where $\kappa=0.4$ is the von K{\`a}rm{\`a}n constant.

\begin{figure*} 
\centering 
  \includegraphics[width=\textwidth]{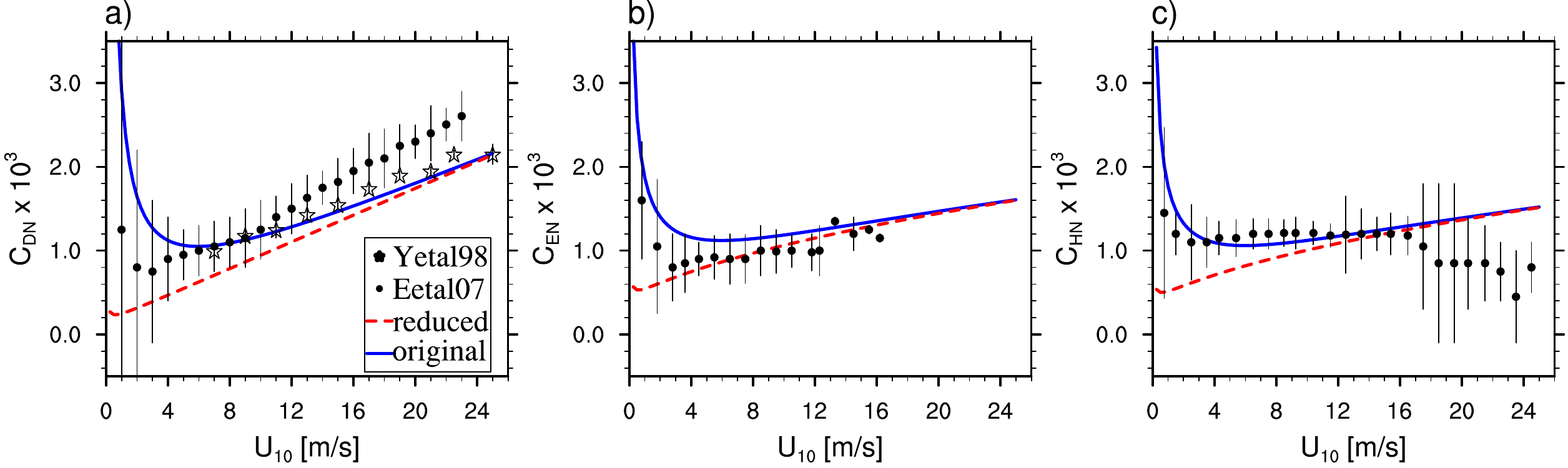}
\caption{Original (solid lines) and reduced (dashed lines)
  neutral drag coefficients for momentum (a), evaporation (b) and
  sensible heat (c) at the air-sea interface as a function of 10\,m
  wind speed in CAM3. Observations for momentum coefficient from
  \cite{Yelland98} (star symbols) and binned data from \cite{Edson08}
  (dots) are overlaid on top. Error bars for binned data denote
  one~standard deviation. Note the large observational uncertainties
  in the low flow speed region.}\label{fig1}
\end{figure*}

Here, the interest is in the effect $Z_{0m}$ has on the zonal-mean
circulation. $Z_{0m}$ is a function of 10\,m wind speed $U_{10}$:
\begin{equation}\label{z0m}
Z_{0m} = 10 \exp\left[-\kappa/ \sqrt{c_4/U_{10} + c_5 +c_6
    U_{10}}\right],
\end{equation}
or equivalently,
\begin{equation}\label{cdnCAM}
C_{DN}=c_4/U_{10} + c_5 +c_6 U_{10}.
\end{equation}
In the original CAM3 surface layer parameterization $c_4, c_5$ and
$c_6$ are given in \cite{Large94}: $c_4=0.0027$\,m\,s$^{-1}$, $c_5 =
0.000142$ and $c_6 = 0.0000764$\,m$^{-1}$\,s.  To reduce
atmosphere-ocean coupling at low flow speed, $c_4$ is set to
$0.000027$\,m\,s$^{-1}$ (see Figure\,\ref{fig1}).  $Z_{0e}$ and
$Z_{0h}$ are given in \cite{Large82}: $Z_{0e}=9.5\times 10^{-5}$\,m,
and $Z_{0h}=2.2\times 10^{-9}$\,m for stable ($\zeta>0$) while
$Z_{0h}=4.9\times 10^{-5}$\,m for unstable ($\zeta\leq 0$)
conditions. It is important to note that $C_{\{E,H\}}$ depend on $C_D$
(equation~\ref{transfer2}). Therefore, reducing $Z_{0m}$ does not only
affect $\boldsymbol{\tau}$, but also $E$ and $H$.

As discussed above, the roughness lengths are empirically derived and
many uncertainties exist in their specification -- especially in the
low flow speed region (e.g., see \cite{Edson08} for the more recent
field campaigns measuring exchange coefficients over open
oceans). Models employing surface layer parameterization other than
\cite{Large82} and \cite{Large94} have different formulations for
roughness lengths -- and hence $C_{\{D,E,H\}}$. For example, the
surface layer exchange formulation in GEOS-5 used in
\cite{Garfinkel11,Garfinkel13} employs the \cite{Helfand95} scheme
with a viscous sublayer to treat the transfer of heat and momentum.

Figure\,\ref{fig1} shows the original (solid line) and the
reduced (dashed line) $C_{\{DN,EN,HN\}}$ profiles used in this
study. The observational data for $C_{DN}$ from \cite{Yelland98} and
binned data from \cite{Edson08} for $C_{\{DN,EN,HN\}}$ are also shown
in the figure. As noted in \cite{Garfinkel11}, the CAM3
parameterization underestimates drag for $U_{10}>12$\,m\,s$^{-1}$ in
comparison to the binned data; however, increasing $C_{DN}$ at high
flow speed does not qualitatively change the results presented
here. Moreover, the aim here is not to tune the surface layer
parameterization scheme but to explore the sensitivity of simulated
zonal-mean circulation to plausible changes in surface layer
parameterization parameters.

\subsection{Setup}\label{setup}

For the aquaplanet setup, the Aqua Planet Experiment (APE) protocol
(http://climate.ncas.ac.uk/ape/) of \cite{Neale00} is followed. In
this setup, the lower boundary is a water-covered surface with no
orography or ice and with a specified zonally-symmetric,
time-invariant SST distribution. The radiative forcing is fixed to
equinoctial insolation and a zonally- and hemispherically-symmetric
ozone distribution is specified. Simulations are forced mostly by the
``Qobs'' SST profile (see figure 1$a$ in \cite{Neale00}), but other
SST profiles, producing different basic states, are also tested in
section~\ref{SSTs} to assess the robustness of the results. The
simulations are run for 36~months, started from a state taken from a
previous aquaplanet simulation. The first 6~months are disregarded as
a spin-up period and the last 30~months of the simulation are averaged
to obtain the climatology. Note that the `true' solution for this
setup is unknown; for example, a broad range of ITCZs across different
models is produced (see \cite{WilliamsonAPE13} figure~3 and
\cite{BlackburnAPE13} figure~4).  In addition, the location and
strength of the ITCZ in aquaplanet setup with CAM3 is resolution and
timestep size sensitive \citep{Williamson08,Williamson13}. However,
the circulation response to reduction in $Z_{0m}$ is unchanged with
different modelling choices.

AMIP-type simulations with observed SSTs (henceforth
AMIPsst) and with a mixed layer motionless slab ocean lower boundary
condition (henceforth AMIPsom) are also discussed. The AMIP-type setup
includes the full complexity and temporal variability of the
underlying surface distribution and seasonal cycle. The AMIPsst setup
is forced by climatological monthly SST and sea-ice distributions. The
AMIPsst simulations are run for ten years and the first year is
disregarded as a spin-up period. In the AMIPsom simulations, annually
averaged mixed layer depths from \cite{MontereyLevitus} are
prescribed. Monthly mean ocean internal heat fluxes are calculated
from the AMIPsst simulations with the original $Z_{0m}$. In the
AMIPsom setup SSTs, ice fraction and ice thickness are predicted. The
simulations are run at T42 horizontal resolution for 100~years and the
last 50~years are used for averaging. The use of the slab ocean
boundary condition guarantees that the surface energy budget is
balanced and that SSTs are free to respond to differences in
parameterization specifications \citep[see discussion in][]{Lee08}.
\begin{figure*}
\centering 
\includegraphics[width=\textwidth]{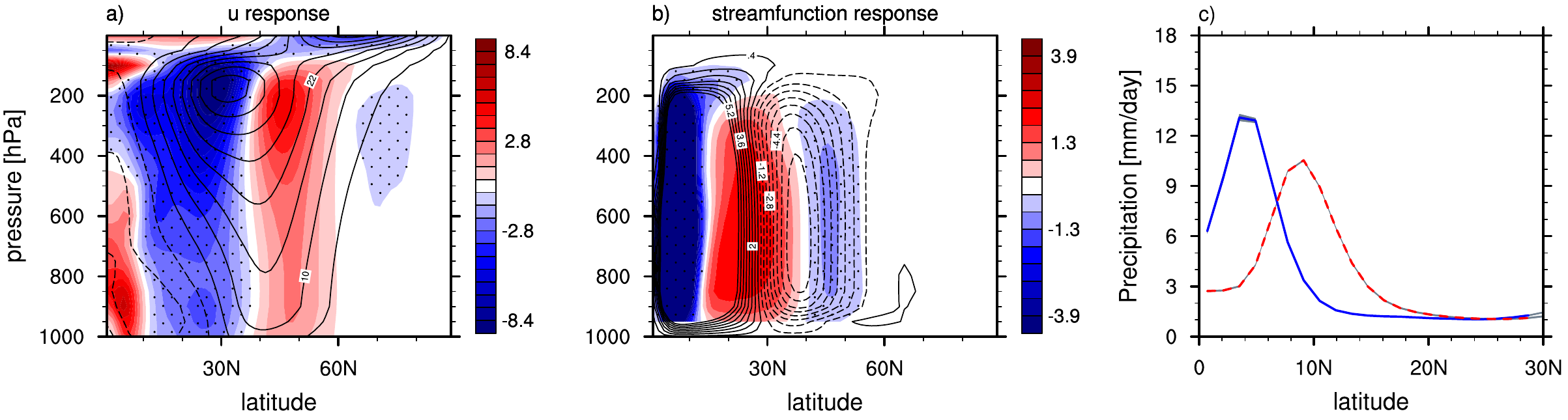}
\caption{Zonal-mean zonal wind $\overline{u}$ [m\,s$^{-1}$] ($a$) and
  Eulerian mass streamfunction $\Psi\times 10^{11}$\,[kg\,s$^{-1}$]
  ($b$) response to reduced $Z_{0m}$ (i.e., reduced $Z_{0m}$ minus the
  original $Z_{0m}$ climatology) in aquaplanet simulations, in
  coloured shading (negative values are stippled).  Contours show the
  original climatology ($6$~m~s$^{-1}$ and 1.6\,kg\,s$^{-1}$ spacing
  for $\overline{u}$ and $\Psi\times 10^{11}$, respectively). The
  negative contours are dashed. Total precipitation [mm\,day$^{-1}$]
  for the original (solid line) and the reduced (dashed line) $Z_{0m}$
  simulations is shown in ($c$). The gray area around the total
  precipitation lines shows the 95\% confidence
  interval. Note that the gray area is barely visible
    because the response is strong and highly
    significant.}\label{fig2}
\end{figure*}

To delineate the impact of reduced $Z_{0m}$ on the large-scale
circulation, all simulations are performed with both the original and
the reduced $Z_{0m}$ (Figure\,\ref{fig1}) and the climatologies
compared. Throughout this study the circulation response to reduced
$Z_{0m}$ that is significant by the Student's t-test at the 95\% level
is shown in coloured shading and the response that falls outside this
significance level is not contoured.

\section{Results: Aquaplanet simulations}\label{sec3}
Figure\,\ref{fig2} shows the zonal-mean zonal wind $\overline{u}$,
Eulerian mass streamfunction $\Psi$, and precipitation $P$ responses
to reduced $Z_{0m}$ for the aquaplanet simulations.
The figure illustrates the main result of this study: the zonal-mean
circulation is quite sensitive to the reduction in $Z_{0m}$ at low
flow speeds.

\begin{figure}
\centering 
\includegraphics[width=0.4\textwidth]{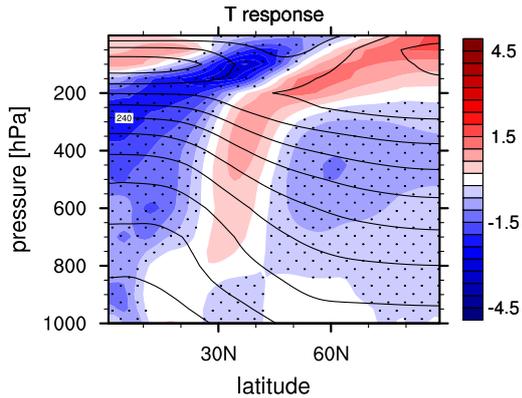}
\caption{Zonal-mean temperature $\overline{T}$ [K] response to reduced
  $Z_{0m}$ in coloured shading (negative values are stippled).
  Contours show the original climatology ($10$~K
  spacing).}\label{fig3}
\end{figure}

Broadly, the response of the zonal-mean circulation to the reduced
$Z_{0m}$ is i) a poleward shift of the mid-latitude westerlies
extending all the way to the surface; ii) a weak poleward shift of the
subtropical descent region; iii) a weakening of the HC (maximum in
$\Psi$ at 500~hPa is $\approx$25\% weaker in the reduced $Z_{0m}$
simulation); and iv) a poleward shift of the tropical upwelling
region, the ITCZ, and the tropical surface easterlies. The zonal-mean
tropospheric temperature $\overline{T}$ response, shown in
Figure~\ref{fig3}, is in thermal wind balance with the mid-latitude
jet response with cooling in the tropical troposphere and polar
regions and warming in the subtropics.

The various aspects of response iv) are related by the atmospheric
Ekman balance -- i.e., that the Coriolis force on the near-surface
meridional mass flux balances the zonal surface wind stress. Responses
i) and ii) are related by the poleward migration of the eddy momentum
flux divergence region (see section~\ref{sec:resp}). Interestingly,
the $\overline{u}$, $\Psi$ and $\overline{T}$ response to reduced
$Z_{0m}$ is similar to the response for La Ni$\tilde{\text{n}}$a minus
El Ni$\tilde{\text{n}}$o phases of ENSO variability (cf.
Figure~\ref{fig2}~and~\ref{fig3} with figure~2~and~4 in \cite{Lu08},
see also \cite{Seager03}). 

It should be noted that the magnitude of the extratropical
$\overline{u}$ response (Figure\,\ref{fig2}a)---especially in the
upper troposphere---is similar to the response to reduced Rayleigh
drag discussed in \cite{Chen07} (their figure~6c). In particular, the
pattern of the response projects strongly onto the annular mode
pattern, defined as the leading EOF of $\overline{u}$ (or surface
pressure). While in \cite{Chen07} the response to the reduced Rayleigh
drag is purely extratropical, here the tropical circulation also
responds to reduced $Z_{0m}$.

The ITCZ---defined as the maximum in $P$---generally coincides with
the zero in $\Psi$, as long as $\Psi$ is not strongly asymmetric about
the ITCZ \citep{Bischoff15}\footnote{The maximum in $P$ generally
  coincides with the HC upwelling region because in steady state $P-E
  = -\partial_y \langle\overline{vq}\rangle \approx -\partial_y
  \langle\bar{v} \bar{q}\rangle\approx q_s \langle\bar{w}\rangle$,
  where $\overline{w}$ is the mean vertical velocity and $E$ does not
  vary strongly with latitude \citep{Numaguti93,Bischoff15}.}. The
poleward shift of the ITCZ in response to the reduced $Z_{0m}$ can
clearly be seen in Figure\,\ref{fig2}$c$.

The response of the ITCZ to the reduced $Z_{0m}$ at low flow speed is
remarkably similar to the response discussed in \cite{Numaguti93} with
fixed $Z_{0m}$\footnote{In his simulations $Z_{0m}=10^{-4}$\,m
  everywhere.}. When he re-introduced the wind dependency to the bulk
formula for $E$ (equation~(\ref{bulks})) instead of fixing $|\Delta
{\bf v}|$=\,6\,m\,s$^{-1}$, the total precipitation transitioned from
having a single peak at the equator to having a double peak straddling
the equator. The double ITCZ formed because $|\Delta {\bf v}|\approx
0$ at the equator implied that $E\approx 0$ there.

\subsection{Other SST profiles}\label{SSTs} 
\begin{figure*}
\centering
$\begin{array}{ccc}
  \includegraphics[width=\textwidth]{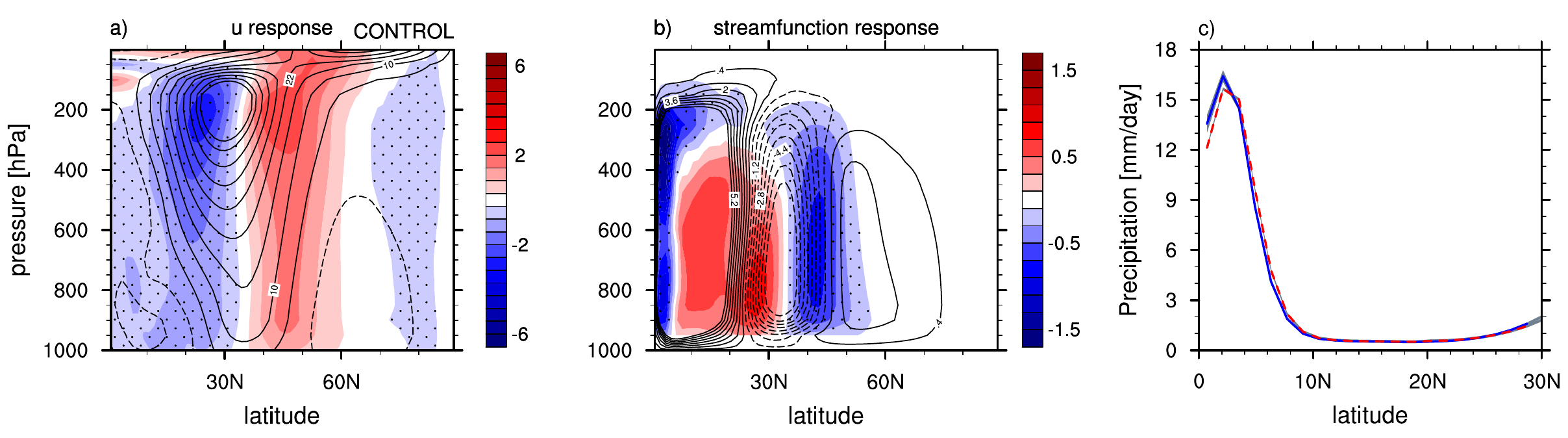}\\
  \includegraphics[width=\textwidth]{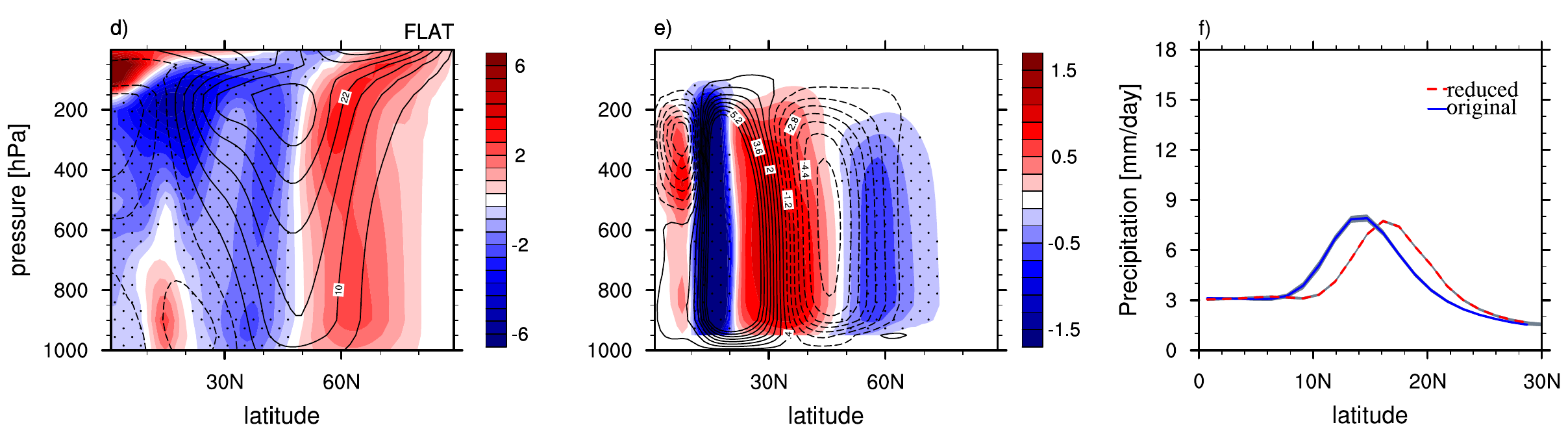}\\
  \includegraphics[width=\textwidth]{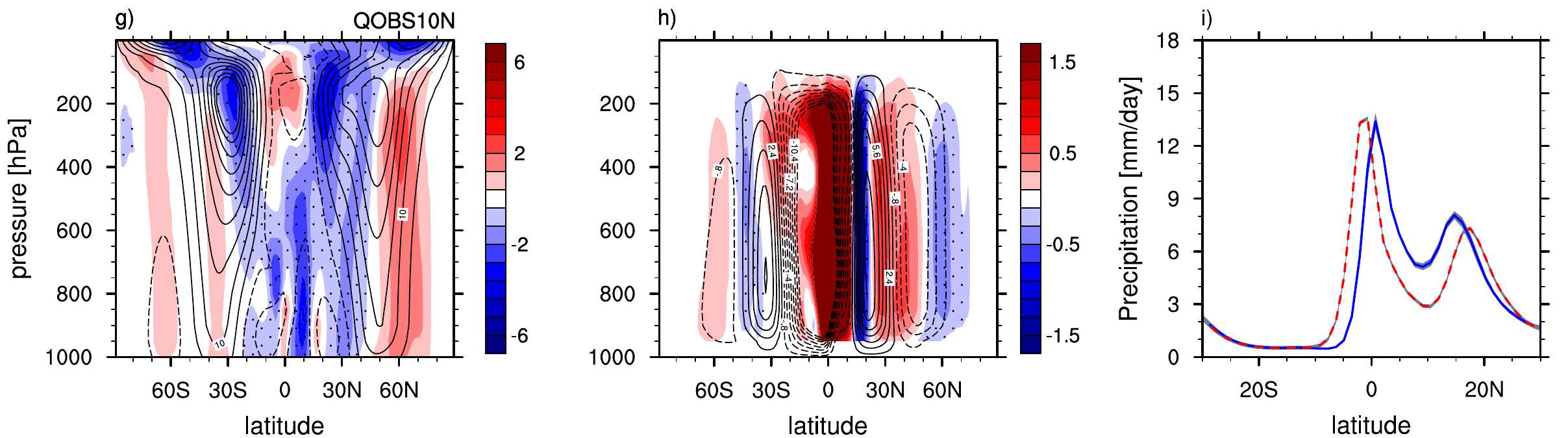}
\end{array}$
\caption{Same as Figure\,\ref{fig2} but for the
    response to reduced $Z_{0m}$ in simulations forced by ``Control''
    (top), ``Flat'' (middle) and ``Qobs10N'' (bottom) SSTs. Note the
    different contour interval than in Figure\,\ref{fig2}. In the
    ``Qobs10N'' case, the latitudinal range in the panels is centred
    at the equator.}\label{fig8new}
\end{figure*}

To check the robustness of the circulation response to reduced
$Z_{0m}$, simulations forced by hemispherically-symmetric ``Control''
and ``Flat'' (figure 1$a$ in \cite{Neale00}) and
hemispherically-asymmetric ``Qobs10N'' (as ``Qobs'' but with maximum
in SST displaced $10\degree$N) SSTs are also performed. The
circulation responses are shown in Figure~\ref{fig8new}. The different
SST profiles produce distinct basic states, especially in the tropics
(see discussion in \cite{WilliamsonAPE13}). Despite the magnitude
being smaller (notice the smaller contour interval), it is clear from
the figure that the responses are similar to that in
Figure~\ref{fig2}. Experience has shown that even under other forcings
(i.e., other than the reduction in $Z_{0m}$), simulations forced by
``Qobs'' SST tend to exhibit a larger magnitude response compared to
the simulations forced by the ``Control'' or ``Flat'' SSTs, which
produce stronger and weaker HC, respectively. The reasons behind this
will be discussed in a future paper. In what follows, the circulation
response is elucidated only in the simulations forced by the ``Qobs''
SST profile.

Note that in the ``Qobs10N'' simulation, the HC return flow rises
above the boundary layer at the equator and gradually crosses to the
hemisphere with the SST maximum in the free troposphere
(Figure~\ref{fig8new}$h$). This also occurs in the AMIP-type
simulations (see Figures\,\ref{fig14}\,and\,\ref{fig15}). Such
behaviour can be associated with too weak a temperature gradient or
too shallow a boundary layer depth at the equator
\citep{Pauluis04}. Consequently, two precipitation maxima are
produced: one almost at the equator, where $\overline{v}$ changes sign
in the lower troposphere, and the other poleward of the SST maximum
where $\overline{v}$ changes sign in the upper troposphere.

\subsection{Tropical energy fluxes as mediators of the response}\label{sec3.3}
Reducing $Z_{0m}$ affects both the tropical and the extratropical
circulation (Figure\,\ref{fig2}). Therefore it is natural to ask if
the zonal-mean circulation response arises from reducing $Z_{0m}$ in
the tropics only, the extratropics only, or both. Because reducing
$Z_{0m}$ affects both latent and sensible heat fluxes ($E$ and $H$,
respectively) and surface stresses ($\boldsymbol{\tau}$)
(equations~\ref{bulks}-\ref{z0m}), it is also natural to ask if the
response is mediated through reduction of $Z_{0m}$ in $H$ and $E$ or
in $\boldsymbol{\tau}$. In other words, is the response
thermodynamically or dynamically initiated?

To address the first question, two aquaplanet simulations are
performed where the reduction in $Z_{0m}$ is
restricted only to latitudes ($\phi$) i) $|\phi| \leq \pm 20\degree$
and ii) $|\phi| > \pm 20\degree$. To address the second question, two
additional simulations are performed where reduction in $Z_{0m}$ is
applied only in the bulk formulation (equation~\ref{bulks}) for i) $H$
and $E$ and ii) $\boldsymbol{ \tau}$.

\begin{figure*}
\centering 
  \includegraphics[width=1.05\textwidth,height=5.2cm]{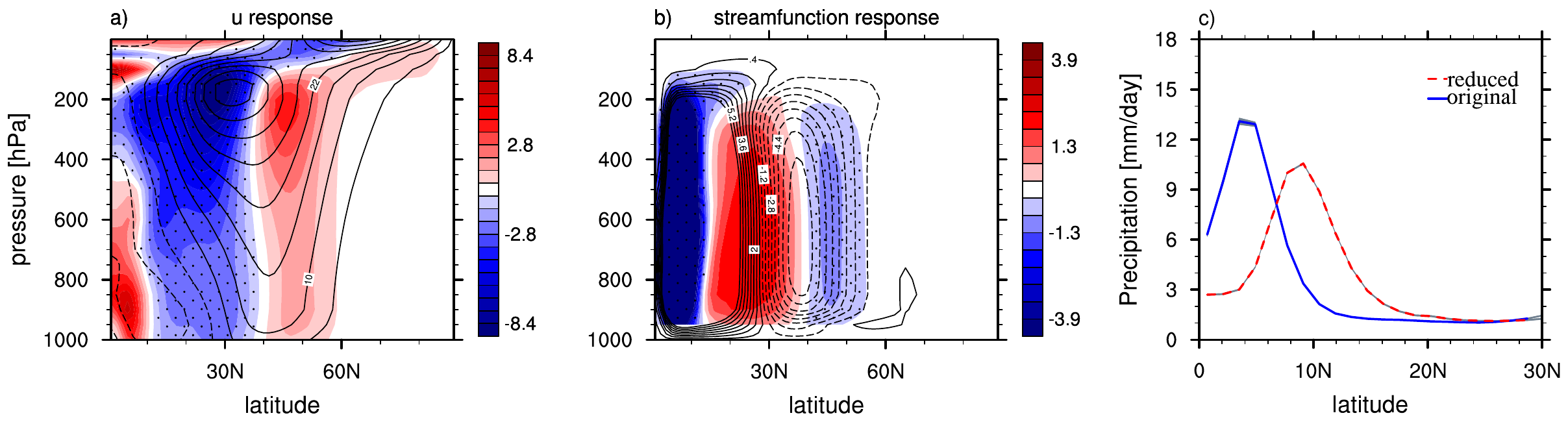}
\caption{Same as Figure\,\ref{fig2} but for the response to reduced
  $Z_{0m}$ in the tropics only, for $|\phi| < 20\degree$.}\label{fig4}
\end{figure*}

\begin{figure*}[!ht]
\centering
  \includegraphics[width=1.05\textwidth,height=5.2cm]{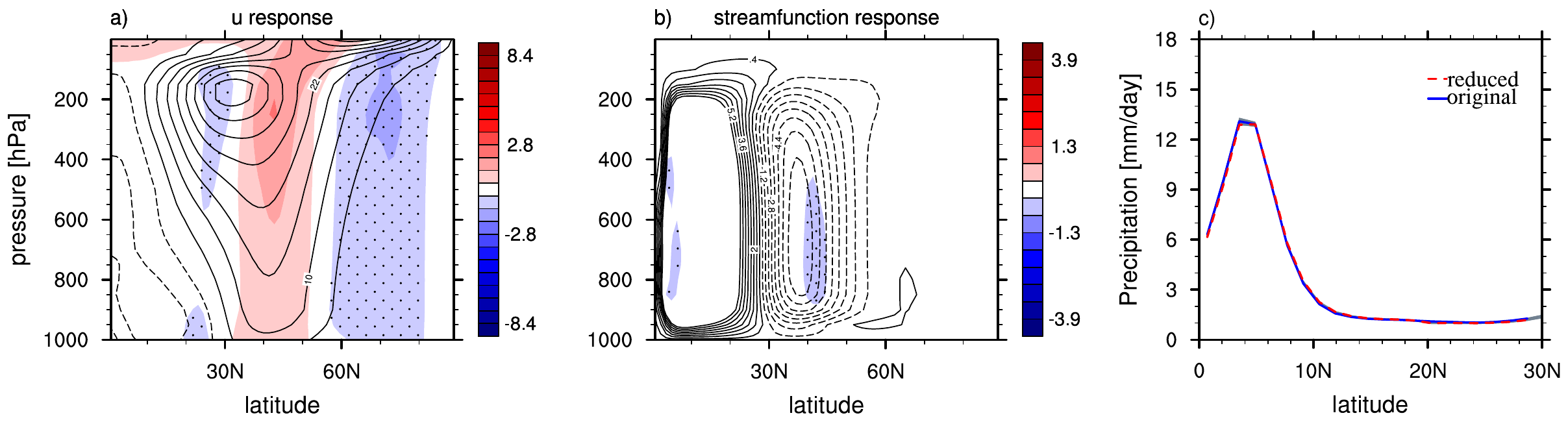}
\caption{Same as Figure\,\ref{fig2} but for the response to reduced
  $Z_{0m}$ in the extratropics only, for $|\phi| >
  20\degree$.}\label{fig5}
\end{figure*}

Figures~\ref{fig4}~and~\ref{fig5} show the zonal-mean circulation
response from the first set of simulations.  By comparing the two
figures with Figure\,\ref{fig2}, it is clear that the response---even
in the mid-latitudes---is mediated from the tropics. This suggests the
existence of a zonally-symmetric tropical-extratropical
teleconnection, which is discussed in section~\ref{sec:resp}.

\begin{figure*} [!ht]
\centering 
  \includegraphics[width=1.05\textwidth,height=5.2cm]{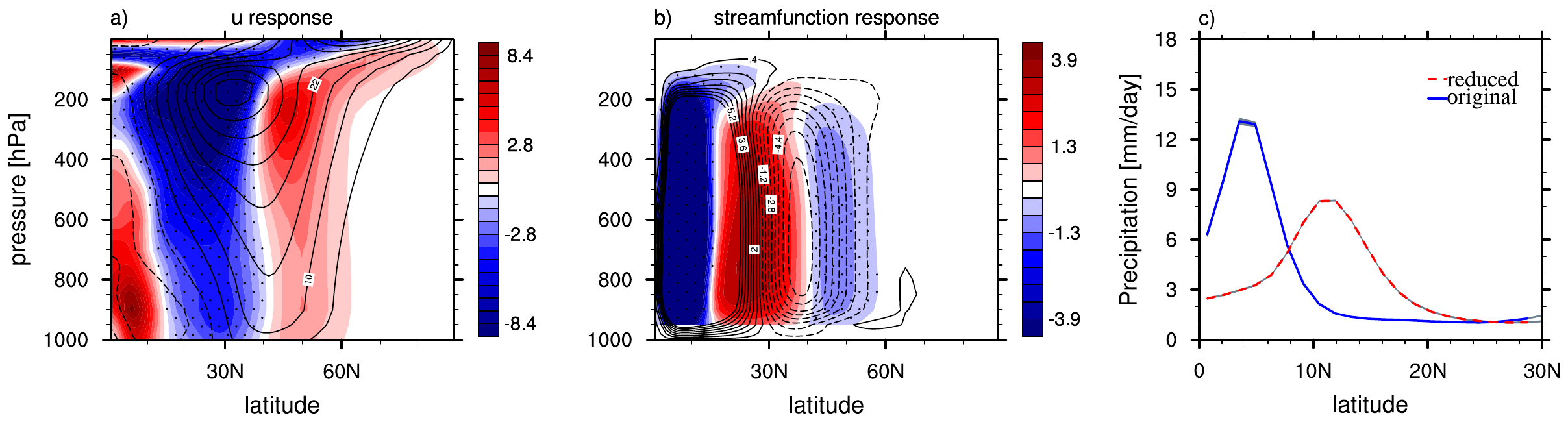} 
\caption{Same as Figure\,\ref{fig2} but for the response to reduced
  $Z_{0m}$ in the bulk formulation for latent and sensible heat fluxes
  only. }\label{fig6}
\end{figure*}

\begin{figure*}[!ht]
\centering 
  \includegraphics[width=1.05\textwidth,height=5.2cm]{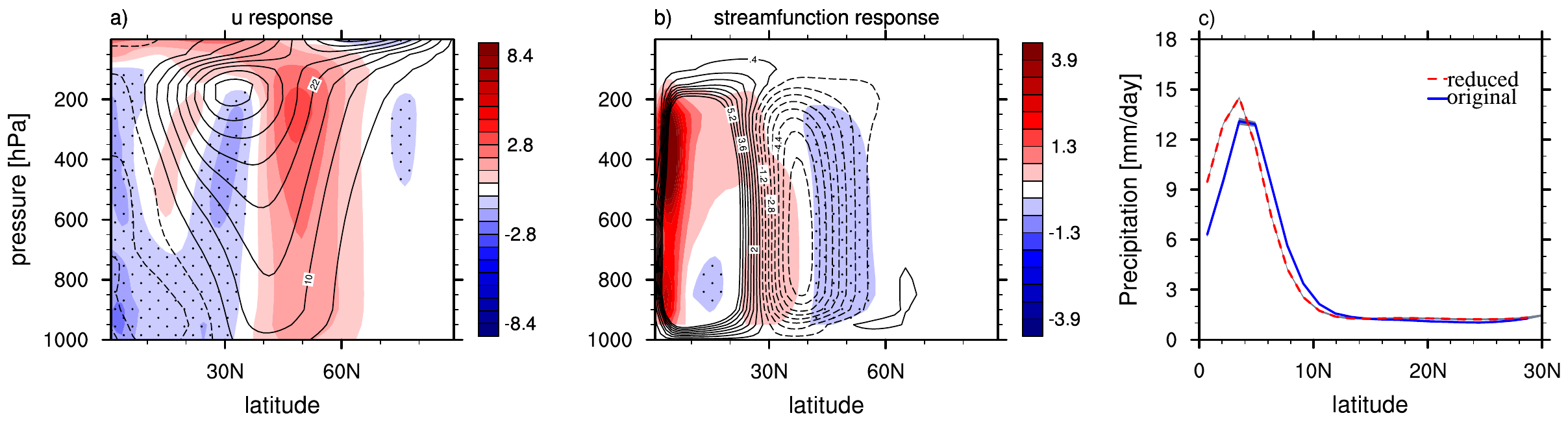} 
\caption{Same as Figure\,\ref{fig2} but for the response to reduced
  $Z_{0m}$ in the bulk formulation for surface stress
  only. }\label{fig7}
\end{figure*}

Figures~\ref{fig6}~and~\ref{fig7} show the zonal-mean circulation
response from the second set of simulations.  By comparing the two
figures with Figure\,\ref{fig2}, it is clear that the total response
is mostly thermodynamically mediated, i.e., through reducing $Z_{0m}$
in $H$ and $E$. Notably, the tropical circulation response to reducing
$Z_{0m}$ in $\boldsymbol{\tau}$ is opposite to the total response; the
HC upwelling region, the ITCZ, and the subtropical jet shift
equatorward in this case. The equatorward shift of the ITCZ in
response to reduction of $\boldsymbol{\tau}$ is a result of the
wind-evaporation feedback \citep[e.g.][]{Neelin87b}: The immediate
response to reducing $\boldsymbol{\tau}$ at low flow speed is to
accelerate the tropical near-surface winds. Because $E$ depends on the
near-surface wind magnitude (equation~\ref{bulks}), the increase in
the tropical surface winds leads to the enhanced $E$ there. As a
result, the HC strengthens (maximum in $\Psi$ at 500~hPa is
$\approx$8\% stronger), the ITCZ shifts equatorward (see
section~\ref{sec:resp} for explanation) and the subtropical jet
becomes stronger (region of positive anomaly in the upper troposphere
equatorward of 20$\degree$N in Figure~\ref{fig7}$a$). This is similar
to the simulation in \cite{Numaguti93}, in which the equatorial flow
speed was artificially enhanced by setting $\Delta {\bf v}
=6$\,m\,s$^{-1}$ everywhere. However, the wind-evaporation feedback is
weak in the simulations shown in Figure~\ref{fig2} because any
increase in wind from reducing $Z_{0m}$ in $\boldsymbol{\tau}$ is
compensated by the reduction of $Z_{0m}$ in the equation that
prescribes $E$.

Note that the ITCZ is displaced further poleward in the simulation
where $Z_{0m}$ is reduced for $E$ and $H$ only
(cf. Figure\,\ref{fig2}$c$ with Figure\,\ref{fig6}$c$). This suggests
that the wind-evaporation feedback acts to partially counteract the
poleward displacement of the ITCZ in the total response
(Figure\,\ref{fig2}).  However, the full response is not linear to the
reduction of $Z_{0m}$ in $\boldsymbol{\tau}$ and in $E$ and $H$ (i.e.,
the sum of the responses in Figures~\ref{fig6}~and~\ref{fig7} is not
the same as the response in Figure~\ref{fig2}). Interestingly, the
mid-latitude westerlies shift poleward in response to reduced
$\boldsymbol{\tau}$ (Figure~\ref{fig7}$a$). Thus it is possible that
the mechanism discussed in \cite{Chen07} that relies on the increase
in wave phase speeds and the resulting poleward shift of the critical
latitude region is playing some role here.


Given that $Z_{0m}$ is reduced at low flow speed only, it is perhaps
not surprising that tropical $E$ and $H$ play a dominant role in the
total response as the extratropical surface winds are directly only
weakly affected. However, in simulations where $Z_{0m}$ is
unrealistically uniformly reduced by a factor of 0.5 to affect all
flow speeds, the response is nearly identical to that in
Figure~\ref{fig2} (not shown). This suggests that the reduction of
$Z_{0m}$ at low flow speed---and especially the effect of this
reduction on $E$ and $H$---is instrumental to the response. When
$Z_{0m}$ is reduced by a factor of 0.5 in the bulk formulation for
$\boldsymbol{\tau}$ only, a stronger extratropical response than in
Figure~\ref{fig7} is produced; the westerlies on the poleward flank of
the mid-latitude jet strengthen and there is an accompanying increase
in wave phase speed (not shown). Thus, if $Z_{0m}$ in
$\boldsymbol{\tau}$ (but only in $\boldsymbol{\tau}$) is substantially
reduced at all flow speeds, the mechanism discussed in \cite{Chen07}
is likely to operate. However, different from \cite{Chen07}, a strong
tropical response is present in that case; the HC strengthens and the
ITCZ shifts equatorward as a result of the wind-evaporation feedback.

Finally, it is worth emphasizing that in simulations forced by other
SST profiles (see Figure~\ref{fig8new}) the tropical energy fluxes are
also instrumental in the total response (not shown).
\begin{figure*}
\centering \includegraphics[width=\textwidth]{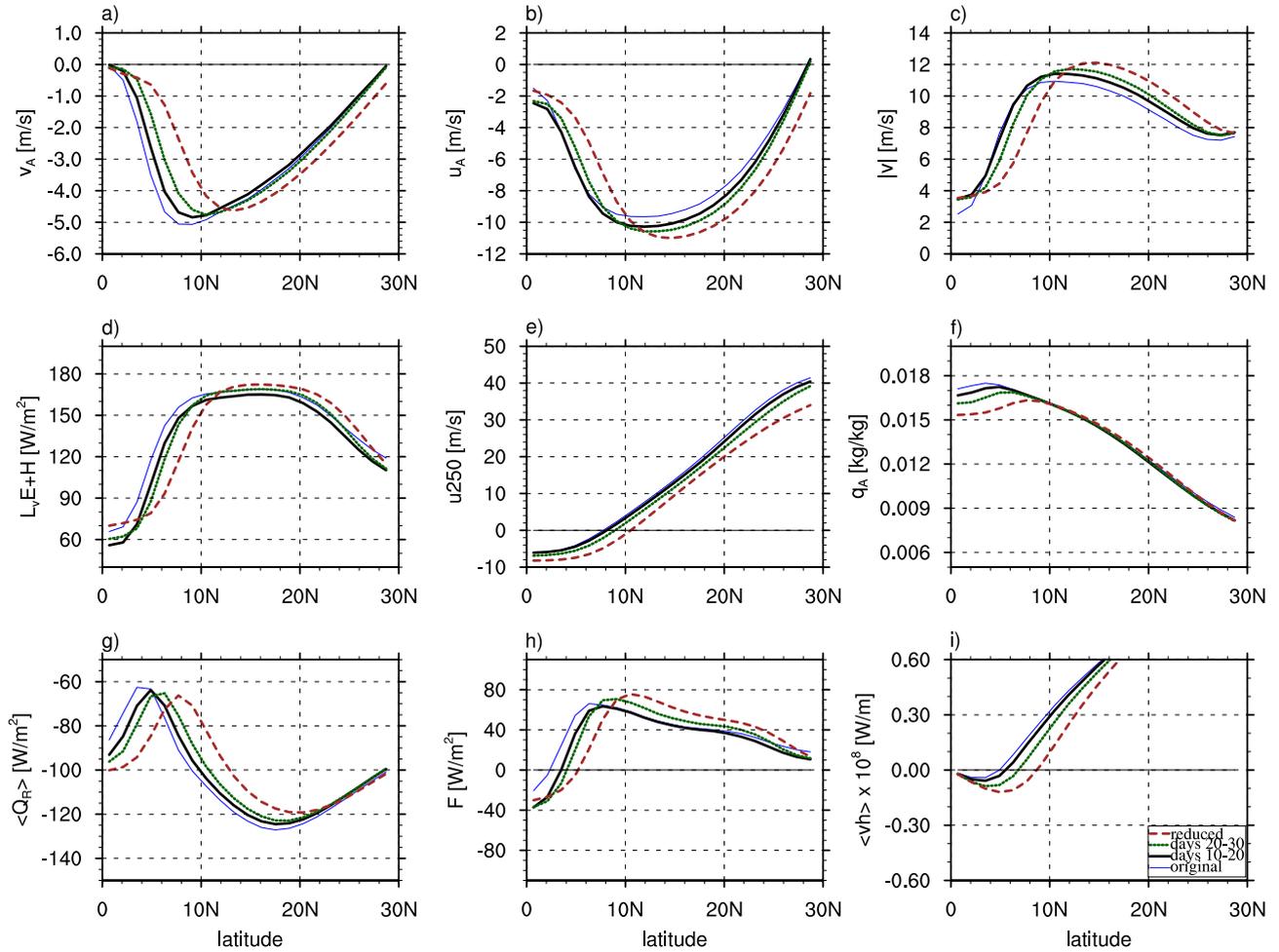}
\caption{(a): The lowest model level meridional wind $v_A$. (b) The
  lowest model level zonal wind $u_A$. (c) $|\Delta {\bf v}| =
  \text{max}(1,\sqrt{u_A^2+v_A^2})$.  (d): $L_v E + H$.  (e): Zonal
  wind at 250\,hPa. (f): The lowest model level specific humidity
  $\overline{q}_A$. (g): $\langle Q_R\rangle$. (h): Net moist static
  energy input ${\cal F}$. (i): Moist static energy flux $\langle
  \overline{vh}\rangle$.  The fields from the equilibrated original
  $Z_{0m}$ simulation are shown in solid lines and from the reduced
  $Z_{0m}$ simulation in dashed lines. The fields averaged over
  $t=$[10:20]\,days are shown in thick lines and over
  $t=$[20:30]\,days in dotted lines.}\label{fig8}
\end{figure*}

\subsection{Understanding the circulation response}\label{sec:resp}
If $Z_{0m}$ is reduced, a potential null hypothesis would be that the
surface winds would accelerate to maintain the same
$\boldsymbol{\tau}$. The increase in surface winds would counteract
the decrease in $C_E$ and $C_H$ leaving $E$ and $H$ unchanged (see
equation~(\ref{bulks})). However, the null hypothesis is clearly
invalid in this study, where the reduction in tropical $E$ and $H$ is
crucial in initiating the response. More importantly, the acceleration
of meridional surface winds implied by the null hypothesis is
inconsistent with Ekman balance, because unchanged $\boldsymbol{\tau}$
implies unchanged Ekman pumping (i.e. in a zonally-symmetric case and
employing an equatorial $\beta$-plane approximation the Ekman pumping
$w_E\sim-\frac{\partial \rho \overline{v_E}}{\partial y} \sim
-\frac{1}{\beta y}\frac{\partial \overline{\tau}_x}{\partial y}$,
where $v_E$ is the meridional wind in the Ekman layer) and hence
unchanged meridional winds in the boundary layer. Therefore, something
else has to happen in the reduced $Z_{0m}$ simulations.

 To examine how the circulation response is established, a 50 ensemble
 member switch-on simulation is performed. In the simulation, the
 reduced $Z_{0m}$ is suddenly switched on in the bulk formulation for
 all the fluxes at $t=10$\,days.  Each member is initialized from a
 random point in the equilibrated original $Z_{0m}$ simulation.
 Figure\,\ref{fig8} shows tropical zonal-mean distributions of fields,
 important for understanding the adjustment, at different stages after
 the switch-on. 

When the reduced $Z_{0m}$ is switched on, the surface meridional wind
($v_A$) weakens initially in the tropics (cf. solid line with thick
line in panel $a$) as a result of the weaker Ekman drag from
$\tau_{x}$ that drives $v_A$. The initial decrease in $\tau_{x}$ can
be seen in Figure\,\ref{fig10new}$a$ (cf. solid line with dashed
line).  The reduction in $\tau_{x}$ is evidently more important than
the decrease in $\tau_{y}$ and in contrast, the initial response in
surface zonal wind ($u_A$) is a strengthening near the equator and at
$10\degree$N$\,<\phi<20\degree$N (panel $b$). Hence, the initial
response is a reduction in $L_vE+H$ everywhere (panel $d$), because
$|\Delta {\bf v}|=\text{max}(1,\sqrt{u_A^2+v_A^2})$ (panel $c$) does
not increase enough to compensate for the decrease in $Z_{0m}$. The
latter is the result of the weakening of $v_A$ just discussed. The
inevitable consequence of reduced $L_vE+H$ is to dry and cool the
tropics.  This can be seen in Figure~\ref{fig3}, where the cooling
(growing with altitude because of the drying) spreads upward in the
deep tropics and poleward in the tropical mid to upper troposphere.

A cooler tropical mid to upper troposphere implies a weakened
subtropical temperature gradient and thus a weakened equatorial flank
of the subtropical jet (panel $e$, also
Figure\,\ref{fig2}$a$)---in analogy with the La Ni$\tilde{\text{n}}$a
phase of ENSO variability (e.g. \cite{Seager03}~and~\cite{Lu08}).
Because the tropical meridional temperature gradient is weaker in the
reduced $Z_{0m}$ simulation, the weaker jet follows from thermal wind
balance. The HC mass flux is also weaker as a weaker HC can now
maintain uniform temperatures in the tropics
\citep[e.g.][]{Schneider06}. Thus another way of understanding the
weakened subtropical jet is weakened angular momentum transport by the
HC. 

\begin{figure}
\centering
  \includegraphics[width=0.45\textwidth]{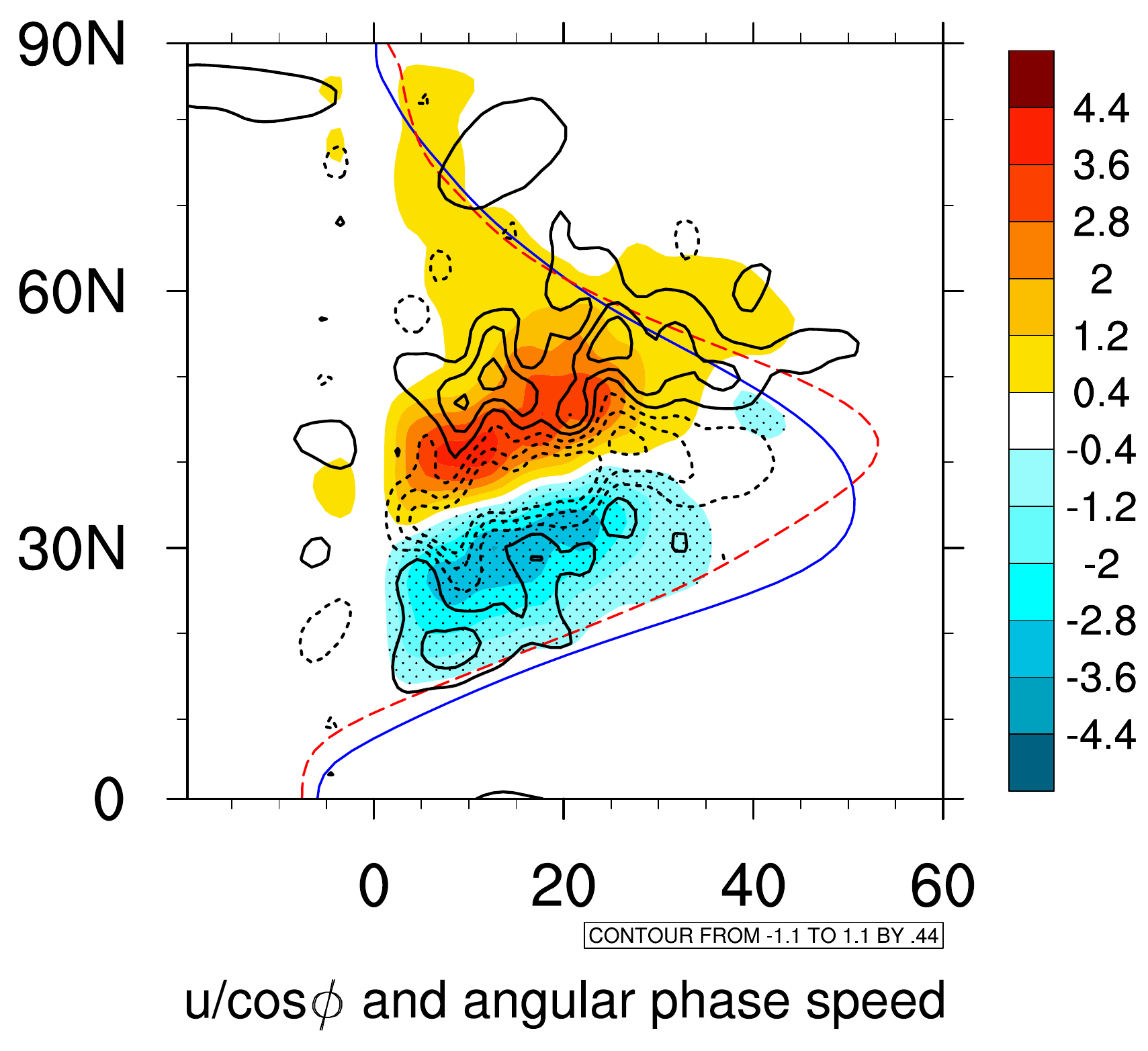}
\caption{Eddy momentum flux convergence spectra $\times
  10^{-7}$\,[m\,s$^{-2}$] at 250\,hPa for the original $Z_{0m}$
  simulation (in coloured shading, negative values are stippled) and
  the response to reduced $Z_{0m}$ (in contours, negative values are
  dashed). The solid and dashed lines show $\overline{u}/\cos\phi$
  [m\,s$^{-1}$] for the original and the reduced $Z_{0m}$ simulations,
  respectively.  The eddy momentum flux cospectrum is calculated
  following \cite{Hayashi71} and \cite{Randel91} before taking the
  divergence.}\label{fig12}
\end{figure}

A weakened equatorial flank of the subtropical jet implies a poleward
shift of the critical latitudes in the subtropics. The critical
latitude is the region where the wave phase speed $c$ closely matches
the mean flow speed $\overline{u}$. It is the region where the Rossby
waves, generated by the near-surface baroclinic instability, break and
deposit easterly momentum \citep{Randel91}. The poleward shift in the
critical latitude in response to reduction in $Z_{0m}$ can be seen in
Figure\,\ref{fig12}, which shows the equilibrated eddy momentum flux
convergence spectra as a function of angular phase speed
($c_A=c/\cos\phi$) and $\phi$ at 250\,hPa.  The figure bears a marked
resemblance to figure~9 in \cite{Lu08} for the El
Ni$\tilde{\text{n}}$o minus La Ni$\tilde{\text{n}}$a ENSO variability,
taking into account the sign reversal. In particular, in the reduced
$Z_{0m}$ simulation there is a systematic poleward shift of the
critical latitude for waves of all $c_A$ and a poleward shift of the
whole eddy momentum flux convergence spectrum, with a characteristic
convergence-divergence-convergence tripole pattern in the response.

It is clear that the weaker subtropical jet leads the poleward
critical latitude shift, because an initial poleward shift of the
critical latitudes would strengthen rather than weaken the winds on
the equatorial flank of the jet (because of less wave drag
there). This poleward critical latitude shift explains the poleward
shift in the HC terminus, and also acts to weaken the core of the jet.

\begin{figure*}
\centering \includegraphics[width=0.9\textwidth]{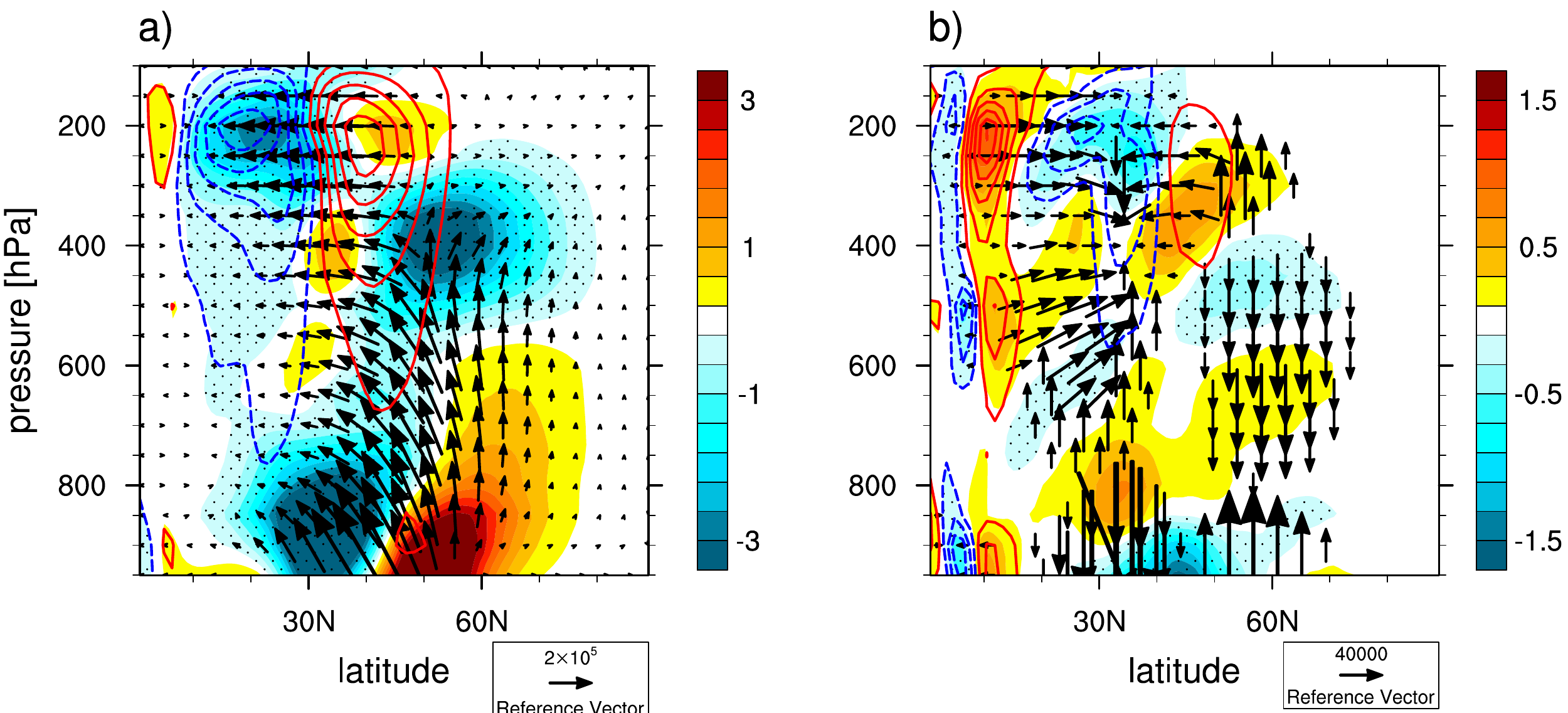}
\caption{Eliassen Palm fluxes [m$^{-2}$\,s$^{-2}$] (vectors), their
  divergence $\times 10^9$ [m\,s$^{-2}$] (in coloured shading,
  negative values are stippled) and eddy momentum flux convergence
  (dashed and solid contours). ($a$) The original $Z_{0m}$
  climatology. ($b$) The response to reduced $Z_{0m}$. Note both the
  equatorward and poleward wave propagation in ($a$).}\label{fig13}
\end{figure*}

\begin{figure*}
\centering \includegraphics[width=\textwidth]{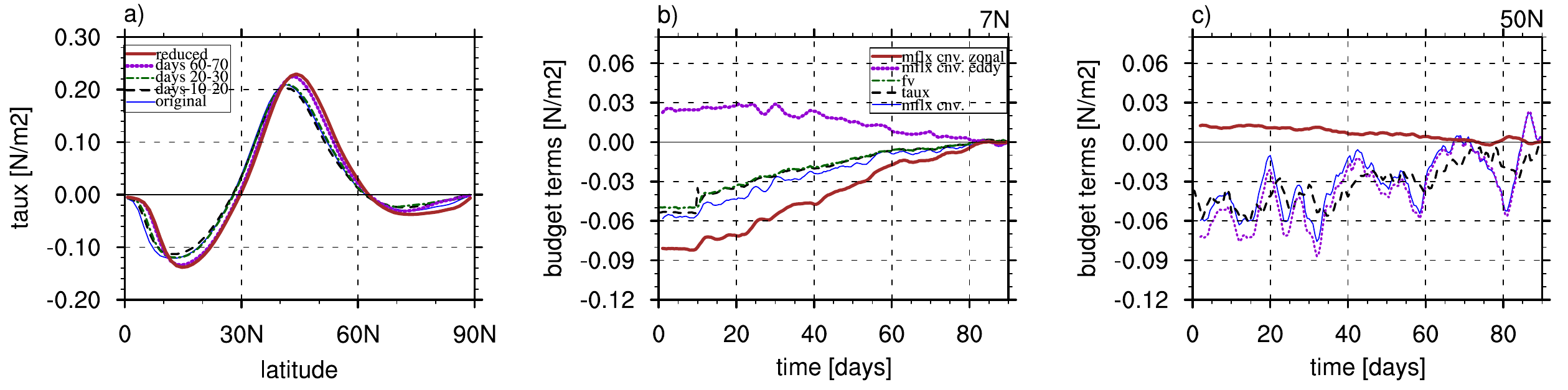}
\caption{(a): The zonal mean zonal surface stress
    $\overline{\tau}_x$ at different stages of the adjustment. (b-c)
    The adjustment of Ekman and momentum balance to the new
    equilibrium obtained for reduced $Z_{0m}$ at (b) 7$\degree$N and
    (c) 50$\degree$N: total meridional momentum flux convergence (thin
    line); eddy meridional momentum flux convergence (dotted line);
    mean meridional momentum flux convergence (thick line); zonal
    surface stress (dashed dotted line); $f\overline{v}_{E}$ (dotted
    line). Panels (b) and (c) show deviations from the final, reduced
    $Z_{0m}$ equilibrium values.}\label{fig10new}
\end{figure*}

Finally, the tropical cooling also reduces baroclinicity in the
subtropical latitudes, reducing the baroclinic eddy generation as
reflected in Figure\,\ref{fig13}, which shows the quasi-geostrophic EP
flux diagnostics \citep{Edmon80} for the original $Z_{0m}$ climatology
in ($a$) and the response in ($b$).  In the original $Z_{0m}$
simulation the EP flux, which represents wave activity propagation,
resembles the classic scenario with upward wave propagation in the
lower and middle troposphere and mostly equatorward wave propagation
aloft. The response in ($b$) shows a reduced EP flux convergence in
mid-latitudes (the red patch in Figure~\ref{fig13}$b$), and thus less
deceleration, which accounts for the strengthened poleward flank of
the subtropical jet. Together with the critical latitude shift
discussed above, this explains the tripole eddy momentum flux
convergence pattern and the poleward jet shift.

To further support the mechanistic chain of events just discussed and
to determine the timescales involved in the adjustment, it is
instructive to examine how the Ekman and zonal momentum
  \footnote{In steady state, vertically integrated meridional momentum
    flux convergence balances $\tau_{x}$.} balances are restored
  following the switch-on.  Figure\,\ref{fig10new} shows the
  adjustment of Ekman and momentum balance terms to their reduced
  $Z_{0m}$ equilibrium counterparts at 7$\degree$N (panel $b$) and at
  50$\degree$N (panel $c$). In order to more easily interpret the
  responses in panels ($b$) and ($c$), the evolution of zonal-mean
  $\tau_x$ during the adjustment is shown for reference in panel
  ($a$). The meridional momentum flux convergence and $f\overline{v}$
  (integrated over the boundary layer) closely follow the spatial
  distribution of $\tau_x$.

After the switch-on, $\tau_{x}$ and $f\overline{v}$ track each other
extremely closely (cf. dotted line with dash dotted line in panels
($b$) and ($c$)), showing that Ekman balance is restored almost
instantly and holds throughout the adjustment process. While the
tropical momentum flux convergence (MFC) also follows $\tau_{x}$
relatively closely, there is a slight imbalance that reflects the lack
of equilibrium (cf. solid line with dotted line). The tropical MFC
adjustment is mainly occurring through the mean MFC term (thick
line). This is a direct consequence of the weakened subtropical
temperature gradient effected by reduced $L_vE+H$. In contrast, the
tropical eddy MFC (dotted line) only starts to respond
$\approx$~15\,days after the switch-on. This shows that the eddy
changes respond to the mean flow changes, rather than vice-versa. In
the extratropics, the MFC and $\tau_{x}$ only start to respond
$\approx$~30 days after the switch-on (panel $c$). In summary, the
upper tropospheric eddy MFC in the subtropics responds to the mean
changes in the tropics and the extratropical changes (reflecting
changes in baroclinic generation) occur later. It is important to
understand that the mechanism behind the poleward midlatitude jet
shift proposed in this study is initiated by the tropical mean flow
changes. In contrast, the mechanism behind the poleward jet shift in
\cite{Chen07} in response to reduced Rayleigh drag is initiated by the
extratropical MFC changes.

\subsection{ITCZ shift}
The arguments in section~\ref{sec:resp} explain the HC weakening and
the poleward shift of the mid-latitude westerlies and the subtropical
descent region. However, the poleward shift of the ITCZ in these
simulations remains to still be elucidated. The moist static energy
budget provides a useful framework for understanding the poleward ITCZ
shift. Neglecting a relatively small contribution from the atmospheric
kinetic energy, the vertically-integrated, zonal-mean moist static
energy budget in a steady state is
\begin{eqnarray}\label{mse}
\partial_{\phi} \langle\overline{vh}\rangle= \langle Q_R\rangle+L_vE +
H \equiv \cal{F},
\end{eqnarray}
where $\langle.\rangle$ denotes a mass-weighted vertical integral over
the atmospheric columns; $\overline{(.)}$ denotes zonal and temporal
mean; $h = c_pT+\Phi+L_v q$ is the moist static energy, where $\Phi$
is the geopotential height, $L_v$ is the latent heat of vaporization,
and $Q_R$ is the net radiative flux. The reader is referred to
\cite{Neelin87a} for the derivation of equation~(\ref{mse}).

Equation~(\ref{mse}) states that the divergence of the moist static
energy flux $ \langle\overline{vh}\rangle$ is balanced by the net
moist static energy input to the atmosphere $\cal{F}$. As long as the
gross moist stability is positive, $\langle\overline{vh}\rangle$ has
the same sign as the mass flux in the upper branch of the HC (see
e.g., discussion in \cite{Bischoff15} in their Appendix 2).
Therefore, the principal quantity controlling the position of the HC
is the net supply of moist static energy, which is a small difference
between the net supply by latent and sensible heat fluxes $L_vE + H$
and the energy loss by radiative cooling $\langle Q_R\rangle$. In
particular, a small fractional change in $L_vE$ and $H$ can cause a
large change in the energy balance of the HC.

As long as the latitude where $\overline{v}$ changes sign does not
vary strongly with height (as in the hemispherically symmetric
simulations in this study), the HC upwelling region and the ITCZ are
located at the latitude where $\langle\overline{vh}\rangle= 0$ ---
i.e., at the ``energy flux equator'' \citep{Kang08,Kang09}. To
understand the ITCZ shift it is instructive to examine the evolution
of $\cal F$ and $\langle \overline{vh} \rangle$ during the adjustment
in Figures\,\ref{fig8}$h$~and~\ref{fig8}$i$. Note that
$\langle\overline{vh} \rangle$ is the latitudinal integral of $\cal
F$, and $\langle\overline{vh} \rangle=0$ at the equator in the
hemispherically-symmetric case discussed here. 

The initial reduction in $L_vE+H$ reduces $\cal F$ in the tropics
(cf. solid line with thick and dotted lines in panel
$h$) and thus increases the latitude at which $\langle
\overline{vh}\rangle$ passes through zero (panel $i$). Thus the ITCZ
is predicted to shift poleward. As the ITCZ shifts poleward, the
surface easterlies and the latitude where $v_A$ changes sign are
pushed poleward to maintain Ekman balance. This leads to a decrease in
$|\overline{\triangle{\bf v}}|$ on either side of the equator
(Figure~\ref{fig8}$c$) and a concomitant decrease in $L_v E + H$ there
(Figure~\ref{fig8}$d$).  Additionally, the minimum in the outgoing
longwave radiation shifts poleward and hence modifies the $\langle Q_R
\rangle$ distribution pushing the maximum in $\langle Q_R \rangle$ at
$5\degree$N poleward (Figure~\ref{fig8}$g$). Hence, the energy flux
equator moves further poleward than initially. As the ITCZ settles to
its final latitude, $\Delta q$ and $\Delta \theta$ begin to increase
to restore $L_vE+H$ ($\Delta q$ increase is reflected in $q$ change in
Figure~\ref{fig8}$f$). Despite this the tropical atmosphere remains
cooler than in the original $Z_{0m}$ simulation as the same flux can
be achieved with a reduced $Z_{0m}$ if the vertical gradient
increases.

\section{Results: AMIPsst and AMIPsom simulations}\label{sec4}
How robust is the zonal-mean circulation response to reduction in
$Z_{0m}$ at low flow speed? To address this, sensitivity experiments
are performed under an AMIP-type setup with both prescribed SSTs (i.e.
AMIPsst) and with a mixed layer slab ocean lower boundary condition
(i.e. AMIPsom)\footnote{Ignoring the sea ice contribution, $T_s$ at
  the lower boundary is governed by
\begin{equation}\label{ssts}
\rho_o c_o h \frac{\partial T_s}{\partial t}=FS-FL-H-L_vE+\cal{Q},
\end{equation}
where $\rho_o$ and $c_o$ are the ocean density and heat capacity,
respectively; $h$ is the annual-mean mixed layer depth; $FS$ is the
net solar flux absorbed by the ocean; $FL$ is the net ocean to
atmosphere longwave flux; and $\cal{Q}$ is the prescribed internal
ocean heat flux convergence.}.  Recall that the AMIP setup includes
the seasonal cycle and the full complexity of
atmosphere-ice-land-ocean interaction.

\begin{figure*}[!ht!]
\centering 
  \includegraphics[width=\textwidth]{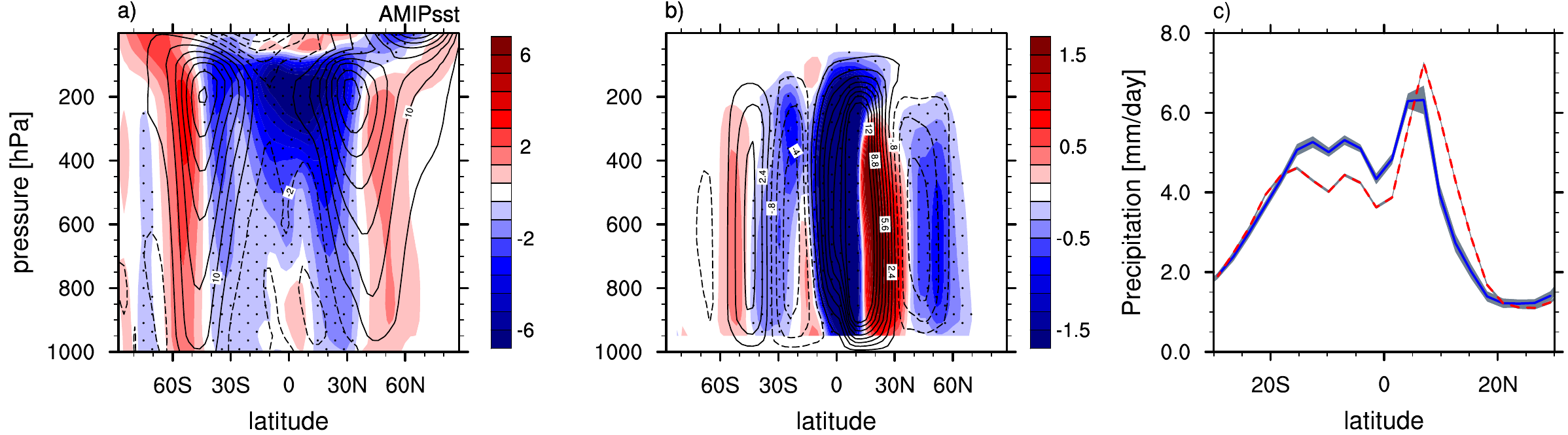}
\caption{Same as Figure\,\ref{fig2} but for the response to reduced
  $Z_{0m}$ in the AMIP-type simulations for DJF with fixed
  SSTs.}\label{fig14}
\end{figure*}

\begin{figure*}[!ht!]
\centering 
  \includegraphics[width=17cm]{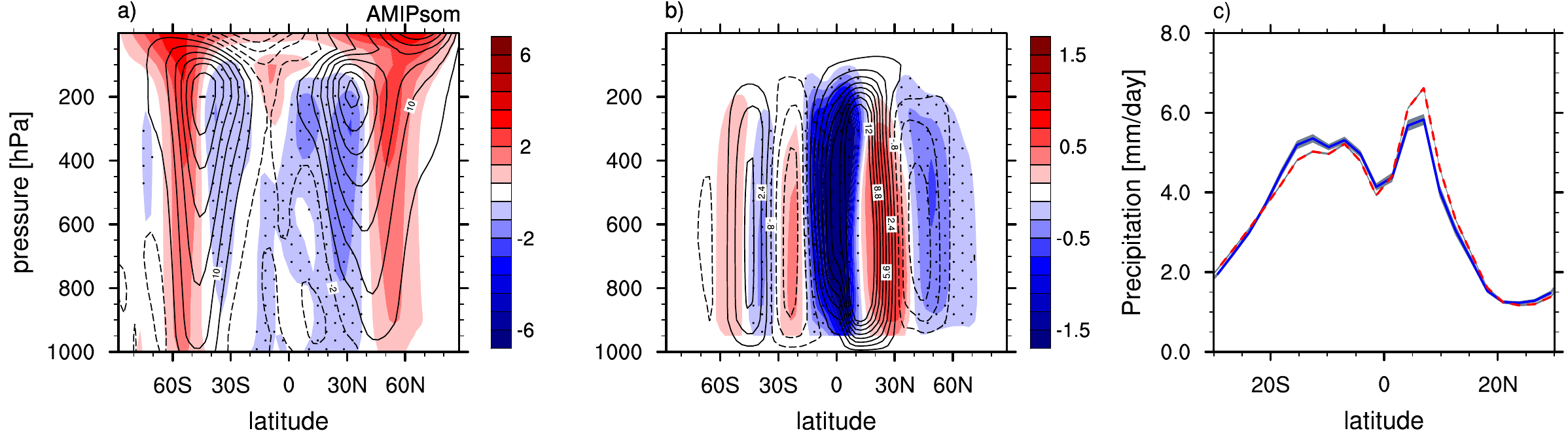}
\caption{Same as Figure\,\ref{fig14} but with slab ocean mixed layer
  lower boundary condition.}\label{fig15}
\end{figure*}

Figures\,\ref{fig14}\,and\,\ref{fig15} show the circulation response
for December-January-February (DJF) to reduced $Z_{0m}$ at low flow
speed in the AMIPsst and AMIPsom simulations, respectively. Similar
responses occur also in other seasons (not shown). Despite the weaker
response compared to the aquaplanet simulations, it is clear from the
figures that the zonal-mean circulation response to reduced $Z_{0m}$
carries over to more complex setups. In particular, the response in
Figures\,\ref{fig14}\,and\,\ref{fig15} closely resembles the response
in the aquaplanet simulation (with hemispheres reversed) forced by the
hemispherically-asymmetric ``Qobs10N'' SST (see Figure~\ref{fig8new},
bottom row). Given that the sensitivity is not substantially
compromised in the presence of the seasonal cycle and the full
complexity of atmosphere-ice-land-ocean interaction, the aquaplanet
framework appears to be a fruitful avenue for understanding the
zonal-mean circulation response in more complex models.

\begin{figure*}[!ht!]
\centering 
  \includegraphics[width=\textwidth,height=4.3cm]{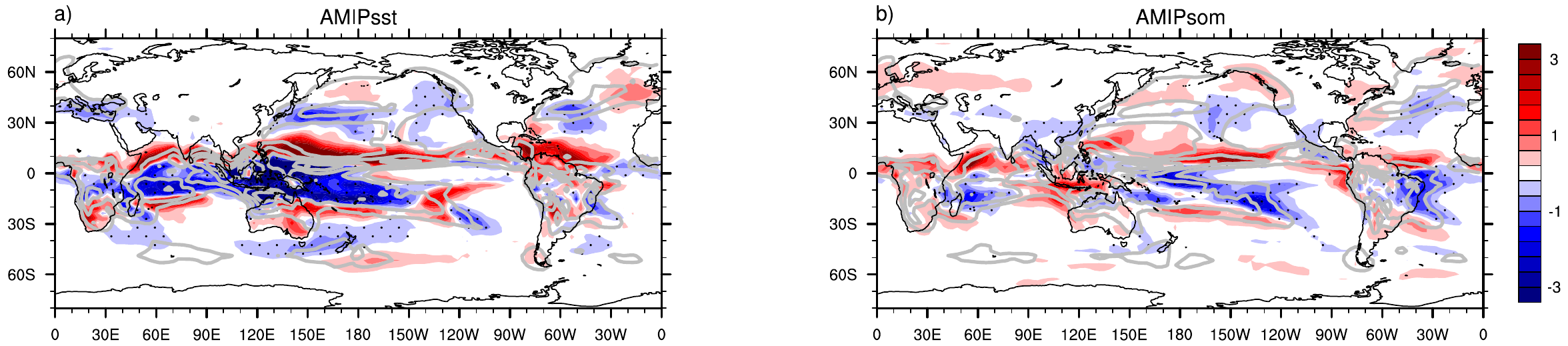}
\caption{Total precipitation response [mm day$^{-1}$] (coloured
  shading, negative values are stippled) for DJF in
  cylindrical-equidistant projection for the AMIP-type simulations
  without ($a$) and with ($b$) a slab ocean mixed layer lower boundary
  condition. The original $Z_{0m}$ climatology is overlaid in gray
  contours.}\label{fig16}
\end{figure*}

Note that the peak in the zonal-mean precipitation maximum in the
summer hemisphere weakens but does not move poleward in response to
reduced $Z_{0m}$ in both experiments. This is because the reduction in
the net equatorial energy input is too small to push the ITCZ
poleward. Note also the presence of a second precipitation maximum in
the winter hemisphere. As discussed earlier, this is because the
latitude where $\overline{v}$ changes sign is not barotropic, with the
HC return flow rising above the boundary layer at the equator and
gradually crossing to the hemisphere with the SST maximum in the free
troposphere (see Figures\ref{fig14}$b$~and~\ref{fig15}$b$).  For a
more detailed view, latitude-longitude cross sections of the total
precipitation response in the AMIPsst and AMIPsom experiments are shown in
Figure~\ref{fig16}. In the figure, the largest precipitation response
occurs over the western and central Pacific and the Indian ocean.


Interestingly, while the circulation sensitivity to reduced $Z_{0m}$
is similar in the AMIPsst and AMIPsom simulations, the global
temperature response is different; the tropospheric temperature warms
globally in the AMIPsom experiments. However, the temperature
response in the slab ocean simulations appears to be rather sensitive
to the internal ocean heat flux and mixed layer depth specification
and is therefore not robust. Detailed investigation into this
difference between the AMIPsst and AMIPsom simulations is, however,
outside the scope of this study.

\section{Summary and Conclusion}\label{sec5}

This study explored the sensitivity of the zonal-mean circulation to
reduction in air-sea momentum roughness $Z_{0m}$ at low flow speed in
CAM3 under three setups: 1) Aquaplanet forced by zonally-symmetric
time-invariant SST, 2) AMIP-type forced by seasonal SST and sea ice
distributions and including the full complexity of
atmosphere-ice-land-ocean interaction, and 3) AMIP-type with a mixed
layer slab ocean lower boundary condition. In all three setups the
circulation response to reduced $Z_{0m}$ at low flow speed is
significant and resembles the La Ni$\tilde{\text{n}}$a minus El
Ni$\tilde{\text{n}}$o differences in ENSO variability
\citep{Seager03,Lu08,SunChenLu13} with; i) a poleward shift of the
mid-latitude westerlies extending all the way to the surface; ii) a
weak poleward shift of the subtropical descent region; iii) a
weakening of the HC, generally also accompanied by a poleward shift of
the ITCZ and the tropical surface easterlies. Mechanism-denial
experiments in the aquaplanet framework pinned down the reduction of
tropical latent and sensible heat fluxes (effected by reducing
$Z_{0m}$ at low flow speed) as the main mediators of the circulation
response. The mechanisms behind the circulation response were also
elucidated in the aquaplanet framework with an ensemble of reduced
$Z_{0m}$ switch-on simulations.

The decrease in the tropical latent and sensible heat fluxes into the
atmosphere cools the tropical troposphere (especially in the mid and
high altitudes) and reduces the meridional temperature gradient there.
As a result the HC weakens and, if the reduction in the energy fluxes
is large enough (as in the aquaplanet simulations with ``Qobs'' SST
profile), the upwelling region and the associated ITCZ shift
poleward. The tropical circulation response is instrumental for the
extratropical circulation response. In particular, a weakened HC is
consistent with a weakened subtropical jet. The weakened jet shifts
the upper tropospheric critical latitude region poleward on the
equatorward flank of the subtropical jet. Importantly, the surface
baroclinicity is also weakened on the equatorward flank of the
midlatitude jet as a result of weaker meridional temperature gradients
in the subtropics. Both the shift in the critical latitude and the
reduced baroclinic eddy forcing result in a tripole EP flux divergence
anomaly in the upper troposphere that induces a poleward shift in the
mid-latitude jet and the HC terminus.

A salient point of this study is that relatively small changes in the
parameterization parameters, within the observational constraints, can
lead to a large response in the circulation. Here only surface layer
parameter sensitivity was explored and in one GCM only. It is possible
that similar changes to the surface layer scheme in GCMs other than
CAM3 produce a different circulation response. For example, the
interaction with other parameterizations might be different;
\cite{Numaguti93} found that the ITCZ moved equatorward (rather than
poleward) in response to reduced (and negative) net moist static
energy input to the equator, because the equatorward transport of the
latent energy surpassed the poleward transport of the dry static
energy (resulting in negative gross moist stability in the tropics) in
the simulations with a dry convective adjustment scheme.  Thus,
delineating the interaction of surface layer parameterizations with
convection and cloud parameterizations deserves further study. It is
also conceivable that the changes to the momentum budget via reduction
in surface stresses play a stronger role in the total response in
other GCMs; for example, the wind-evaporation feedback potentially
could overcome the weakening of the HC and the poleward ITCZ shift.

As is shown here, uncertainties in surface process parameterizations
are likely to contribute to the systematic errors in GCM simulations
together with cloud and convective parameterizations, which are often
singled out as sole culprits. Furthermore, it is plausible that
reasonable changes in other parameterization parameters can lead to as
large a circulation response as observed here. Therefore, it is
important to understand such parameter sensitivities in order to
delineate non-robust circulation responses to climate forcing
\citep[e.g.][]{Stevens13,Shepherd14}.

\section*{Acknowledgments}
The authors thank Mike Blackburn, Terry Davies, Felix Pithan and Alan
Plumb for useful discussions pertaining to this work. The two
anonymous reviewers are thanked for their useful comments, which
greatly improved the manuscript. This study is supported by the
`Understanding the atmospheric circulation response to climate change'
(ACRCC, ERC Advanced Grant) project.

\end{document}